\documentclass[prl,aps,twocolumn,showpacs,superscriptaddress,floatfix]{revtex4}
\usepackage{amsfonts}
\usepackage{amssymb}
\usepackage{graphicx}
\usepackage{amsmath}
\usepackage{mathptmx}
\usepackage{eucal}

\begin{document}

\title{Enhancement of critical current by
microwave irradiation in wide superconducting films}

\author{V.M.~Dmitriev}
\affiliation{Institute for Low Temperature Physics and
Engineering, Kharkov 61103, Ukraine}
\affiliation{International Laboratory of High Magnetic Fields and
Low Temperatures, Wroclaw 53-421, Poland}

\author{I.V.~Zolochevskii}
\affiliation{Institute for Low Temperature Physics and
Engineering, Kharkov 61103, Ukraine}

\author{E.V.~Bezuglyi}
\affiliation{Institute for Low Temperature Physics and
Engineering, Kharkov 61103, Ukraine}

\begin{abstract}
The temperature dependences of the enhanced critical current in
wide and thin Sn films exposed to the microwave field have been
investigated experimentally and analyzed. It was found that the
microwave field stabilizes the current state of a wide film with
respect to the entry of Abrikosov vortices. The stabilizing effect
of irradiation increases with frequency. Using similarity between
the effects of microwave enhancement of superconductivity observed
for homogeneous (narrow films) and inhomogeneous (wide films)
distributions of the superconducting current over the film width,
we have succeeded in partial extension of the Eliashberg theory to
the case of wide films.
\end{abstract}


\maketitle

\section{Introduction}

During last decades, current states in wide superconducting films
in the absence of external magnetic and microwave fields have been
studied in considerable detail. The main property of wide films,
which distinguishes them from narrow channels, is an inhomogeneous
distribution of the transport current over the film width. This
distribution is characterized by increase in the current density
towards the film edges due to the Meissner screening of the
current-induced magnetic field. It should be emphasized that the
current state of a wide film is qualitatively different from the
Meissner state of a bulk current-carrying superconductor, despite
their seeming resemblance. Indeed, whereas the transport current
in the bulk superconductor flows only within a thin surface layer
and vanishes exponentially at the distance of the London
penetration depth $\lambda(T)$ from the surface, the current in
the wide film is distributed over its width $w$ more uniformly,
according to the approximate power-like law $[x(w-x)]^{-1/2}$
\cite{Larkin,Aslamazov}, where $x$ is the transversal coordinate.
Thus the characteristic length $\lambda_\perp(T)=2\lambda^2(T)/d$
($d$ is the film thickness), which is commonly referred to as the
penetration depth of the perpendicular magnetic field, has nothing
to do with any spatial scale of the current decay with the
distance from the edges. In fact, the quantity $\lambda_\perp(T)$
plays the role of a ``cutoff factor'' in the above-mentioned law
of the current distribution at the distances $x,w-x \sim
\lambda_\perp$ from the film edges and thereby determines the
magnitude of the edge current density. In films whose width is
much larger than $\lambda_\perp(T)$ and the coherence length
$\xi(T)$, the edge current density approaches the value
$j_0=I/d\sqrt{\pi w\lambda_\perp}$ \cite{Larkin} if the total
current $I$ does not exceed the homogeneous pair-breaking current
$I_{\rm c}^{\rm GL}$.

In such an inhomogeneous situation, the mechanism of
superconductivity breaking by the transport current differs from
homogeneous Ginzburg-Landau (GL) pair-breaking in narrow channels.
In wide films this mechanism is associated with disappearance of
the edge barrier for the vortex entry into the film when the
current density at the film edges approaches the value of the
order of the GL pair-breaking current density $j_{\rm c}^{\rm GL}$
\cite{Larkin,Aslamazov,Shmidt,Likharev}. Using a qualitative
estimate of $j_0 \approx j_{\rm c}^{\rm GL}$ for the current
density which suppresses the edge barrier, one thus derives the
expression $I_{\rm c}(T) \approx j_{\rm c}^{\rm GL}(T)d\sqrt{\pi
w\lambda_\perp(T)}$ for the critical current of a wide film. This
equation imposes a linear temperature dependence of the critical
current, $I_{\rm c}(T) \propto (1-T/T_{\rm c})$, near the critical
temperature $T_{\rm c}$, and it is widely used in analysis of
experimental data (see, e.g., \cite{Andratskiy}). A quantitative
theory of the resistive states of wide films by Aslamazov and
Lempitskiy (AL) \cite{Aslamazov} also predicts the linear
temperature dependence of $I_{\rm c}$ but gives the magnitude of
the critical current larger by a factor of $1.5$ than the above
estimate of $I_{\rm c}$. This result was supported by recent
experiments on the critical current in wide films \cite{Dmitriev}.

Since the parameters $\xi(T)$ and $\lambda_\perp(T)$ infinitely
grow while the temperature approaches $T_{\rm c}$, any film
reveals the features of a narrow channel in the immediate vicinity
of $T_{\rm c}$; in particular, its critical current is due to the
uniform pair-breaking (narrow channel regime) thus showing the
temperature dependence of the GL pair-breaking current $I_{\rm
c}^{\rm GL}(T) \propto (1-T/T_{\rm c})^{3/2}$. As the temperature
decreases, the film exhibits a crossover to an essentially
inhomogeneous current state, in which vortex nucleation is
responsible for the resistive transition; in what follows, this
regime will be referred to as a wide film regime. A following
quantitative criterion of such a crossover was formulated in
\cite{Dmitriev} on the basis of careful measurements: if the
temperature $T$ satisfies an implicit condition
$w<4\lambda_\perp(T)$, the superconducting film can be treated as
a narrow channel, whereas at $w>4\lambda_\perp(T)$ it behaves as a
wide film. The physical interpretation of this criterion is quite
simple: existence of the resistive vortex state requires at least
two opposite vortices (vortex and anti-vortex), with the diameter
$2\lambda_\perp$ each, to be placed across the film of the width
$w$. At the same time, it was noted in \cite{Dmitriev} that after
entering the wide film regime, the dependence $I_{\rm c}(T)
\propto (1-T/T_{\rm c})^{3/2}$, typical for narrow channels, still
holds within a rather wide temperature range, though the absolute
value of $I_{\rm c}$ is lower than the pair-breaking current
$I_{\rm c}^{\rm GL}(T)$. In practice, the linear temperature
dependence \cite{Aslamazov} of the critical current becomes
apparent only at low enough temperatures, when $\lambda_\perp(T)$
becomes 10--20 times smaller than the film width.

Whereas the equilibrium critical current of wide films has been
quite well studied, an interesting question about current states
in the films under nonequilibrium conditions was much less
investigated. In recent papers \cite{Agafonov,Dmitriev1} it was
first reported that wide superconducting films, similar to short
bridges \cite{bridges} and narrow channels \cite{narrow},
exhibit increase in the critical current under microwave
irradiation (superconductivity enhancement).
In this paper, we present results of systematic investigations of
the enhanced critical current in wide superconducting films. We
argue that all essential features of the enhancement effect in
wide films with an inhomogeneous current distribution are very
similar to that observed before in narrow channels
\cite{narrow}. We found that microwave irradiation stabilizes
the current state in respect to the vortex nucleation and thus
considerably extends the temperature region of the narrow channel
regime. The relative moderateness of the current inhomogeneity in
wide films enables us to exploit the theory of the
superconductivity stimulation in spatially homogeneous systems
with minor modifications for a quantitative treatment of our
experimental data.

\section{Nonequilibrium critical current of superconducting channels
in microwave field}

The theory of superconductivity enhancement under microwave
irradiation was created by Eliashberg
\cite{Eliashberg1,Eliashberg2,Ivlev} for superconducting systems,
in which the equilibrium energy gap $\Delta$ and the
superconducting current density $j_{\rm s}$ are distributed
homogeneously over the sample cross-section. The theory applies to
rather narrow and thin films [$w,d\ll\xi(T),\lambda_\perp(T)$]
with the homogeneous spatial distribution of the microwave power
and, correspondingly, of the enhanced gap. The length of electron
scattering by impurities, $l_{\rm i}$, is assumed to be small
compared to the coherence length. According to this theory, the
effect of the microwave irradiation on the energy gap $\Delta$ of
a superconductor carrying the transport current with the density
$j_{\rm s}$ is described by the generalized GL equation,
\begin{eqnarray} \label{GLnoneq}
\frac{T_{\rm c} - T}{T_{\rm c}}-\frac{7\zeta(3)\Delta^2}{8(\pi k
T_{\rm c})^2}-\frac{2kT_{\rm c}\hbar}{\pi e^2 D \Delta^4
N^2(0)}j_{\rm s}^2 +\Phi(\Delta)=0.
\end{eqnarray}
Here $N(0)$ is the density of states at the Fermi level, $D=v_{\rm
F} l_{\rm i}/3$ is the diffusion coefficient, $v_{\rm F}$ is the
Fermi velocity, and $\Phi(\Delta)$ is a nonequilibrium term
arising from the nonequilibrium addition to the electron
distribution function \cite{Eliashberg1,Eliashberg2,Klapwijk},
\begin{align}
\Phi(\Delta)&=-\frac{\pi \alpha}{2kT_{\rm c}}\left[
1+0.11\frac{(\hbar \omega)^2}{\gamma kT_{\rm c}} -\frac{(\hbar
\omega)^2}{2\pi \gamma \Delta}\left(\ln \frac{8 \Delta}{\hbar
\omega}-1\right)\right],\nonumber \\
 \hbar \omega &< \Delta.\label{Phi}
\end{align}
In this equation, $\alpha=Dp_{\rm s}^2/ \hbar$ is the quantity
proportional to the irradiation power $P$, $p_{\rm s}$ is the
amplitude of the superfluid momentum excited by the microwave
field, $\gamma = \hbar/ \tau_\varepsilon$, and $\tau_\varepsilon$
is the energy relaxation time.

In our studies of the superconductivity enhancement, we usually
measure the critical current rather than the energy gap. Thus, in
order to compare our experimental data with the Eliashberg theory,
we should express the superconducting current density $j_{\rm s}$
through the function of the energy gap, temperature and
irradiation power, using \eqref{GLnoneq} and \eqref{Phi},

\noindent
\begin{align} \label{js}
j_{\rm s}&= \eta \Delta^2 \left[ \frac{T_{\rm c} - T}{T_{\rm
c}}-\frac{7\zeta(3)\Delta^2}{8(\pi k T_{\rm c})^2}+\Phi(\Delta)
\right] ^{1/2},
\\
\eta &=eN(0) \sqrt\frac{\pi D}{2 \hbar kT_{\rm c}}.\label{eta}
\end{align}
The extremum condition for the superconducting current, $\partial
j_{\rm s}/\partial \Delta=0$, at given temperature and irradiation
power results in a transcendental equation for the energy gap
$\Delta$,
\begin{align}
\frac{T_{\rm c}-T}{T_{\rm c}}&-\frac{21\zeta(3)\Delta^2}{(4\pi k
T_{\rm c})^2}-\frac{\pi \alpha}{2kT_{\rm c}} \left[
1+0.11\frac{(\hbar \omega)^2}{\gamma kT_{\rm c}}\right.\nonumber\\
&- \left.\frac{(\hbar \omega)^2}{4\pi \gamma
\Delta}\left(\frac{3}{2} \ln \frac{8 \Delta}{\hbar
\omega}-1\right) \right] =0.\label{EqDelta}
\end{align}
The solution of this equation, $\Delta=\Delta_{\rm m}$, being
substituted into \eqref{js}, determines the maximum value of
$j_{\rm s}$, i.e., the critical current \cite{Dmitriev2},
\begin{equation} \label{IcP}
I_{\rm c}^{\rm P}(T) =\eta d w \Delta_{\rm m}^2 \left[
\frac{T_{\rm c}-T}{T_{\rm c}}-\frac{7\zeta(3)\Delta_{\rm
m}^2}{8(\pi k T_{\rm c})^2} + \Phi(\Delta_{\rm m}) \right] ^{1/2}.
\end{equation}
This basic equation for the enhanced critical current will be used
throughout this paper. With no microwave field applied
($\alpha=0$), equation \eqref{IcP} transforms into the following
expression for the equilibrium pair-breaking current,
\begin{eqnarray} \label{IcGL}
I_{\rm c}(T)= I_{\rm c}^{\rm GL}(T) = \eta d w \Delta_{\rm m}^2
\left[ \frac{T_{\rm c}-T}{T_{\rm c}}-\frac{7\zeta(3)\Delta_{\rm
m}^2}{8(\pi k T_{\rm c})^2} \right]^{1/2},
\end{eqnarray}
where $\Delta_{\rm m}=\sqrt{2/3} \Delta_0$, and
\begin{equation} \label{Deltaeq}
\Delta_0= \pi k T_{\rm c} \sqrt{8(T_{\rm c}- T)/7\zeta(3)T_{\rm
c}}=3.062 k T_{\rm c} \sqrt{1-T/T_{\rm c}}
\end{equation}
is the equilibrium value of the gap at zero transport current.

When using equation \eqref{eta} for the quantity $\eta$ with the
density of states given by the free electron model,
$N(0)=m^2v_{\rm F}/ \pi^2 \hbar^{3}$, in calculation of the
equilibrium critical current \eqref{IcGL}, we faced with a
considerable discrepancy of \eqref{IcGL} with experimental values
of $I_{\rm c}^{\rm GL}(T)$. This implies that such an estimate of
$N(0)$ for the metal (Sn) used in our experiments is rather rough.
The way to overcome such an inconsistency is to express $N(0)$
through the experimentally measured quantity, viz., film
resistance per square, $R^\square=R_{4.2}w/L$, where $R_{4.2}$ is
the total film resistance at $T=4.2 K$ and $L$ is the film length.
Then equation \eqref{eta} transforms to the relation
\begin{equation}
\eta=(e d R^\square)^{-1} \sqrt{3 \pi /2k T_{\rm c} v_{\rm F}
l_{\rm i} \hbar}, \nonumber
\end{equation}
which provides good agreement of equation \eqref{IcGL} with both
the experimental values of equilibrium pair-breaking current and
the values calculated through the microscopic parameters of the
film, $\xi_0$ and $\lambda_\perp(0)$ [see \eqref{IcGL1} below].

Curiously, as far as we know, the temperature dependences of the
enhanced critical current following from \eqref{IcP} have been
never quantitatively compared with experimental data. However,
qualitative attempts of a comparison of the experimental
dependences $I_{\rm c}^{\rm P}(T)$ with the Eliashberg theory have
been done. For instance, the authors of \cite{Klapwijk}
interpreted their experimental results on the enhancement effect
in narrow films in a following way. First, using the relation
\eqref{Deltaeq} for the equilibrium gap, they presented the GL
pair-breaking current as
\begin{eqnarray} \label{IcGL1}
I_{\rm c}^{\rm GL}(T)=
\frac{c\Phi_0w}{6\sqrt{3}\pi^2\xi(0)\lambda_\perp(0)}(1- T/T_{\rm
c})^{3/2}=K_1 \Delta_0^3(T),
\end{eqnarray}
where $\Phi_0=hc/2e$ is the magnetic flux quantum. Then, using an
empiric fact that the temperature dependence of the enhanced
critical current in the narrow film is close to the equilibrium
one, $I_{\rm c}^{\rm P}(T) \propto (1-T/T_{\rm c}^{\rm P})^{3/2}$,
where $T_{\rm c}^{\rm P}$ is a superconducting transition
temperature in a microwave field, this dependence was modelled by
equation similar to \eqref{IcGL1},
\begin{eqnarray}\label{Klap}
I_{\rm c}^{\rm P}(T)=K_2 \Delta_P^{3}(T).
\end{eqnarray}
The enhanced energy gap $\Delta_P(T)$ in \eqref{Klap} was
calculated by the Eliashberg theory for zero superconducting
current [equation \eqref{GLnoneq} at $j_{\rm s}=0$]. Assuming
$K_1=K_2$ and using the magnitude of the microwave power as a
fitting parameter, the authors of \cite{Klapwijk} eventually
achieved good enough agreement between the calculated and measured
values of $I_{\rm c}^{\rm P}(T)$.

Obviously, such a comparison of the experimental data with the
Eliashberg theory should be considered as a qualitative
approximation which cannot be used to obtain quantitative results.
First, equations \eqref{IcGL1} and \eqref{Klap} involve the gap
value at zero current ($j_{\rm s}=0$) which is different from that
with the current applied. Second, the pair-breaking curve $j_{\rm
s}(\Delta)$ in the equilibrium state, which is implicitly assumed
in \eqref{Klap}, significantly differs from that in the microwave
field \cite{Dmitriev2}. Of course, such model assumptions might
nevertheless give a relatively good numerical approximation to a
right formula (actually, that is what has been used in
\cite{Klapwijk}), yet the basic inconsistency of such an
approximation is due to qualitatively different behaviour of
$I_{\rm c}^{\rm P}(T)$ and $\Delta_P(T)$ in the vicinity of the
critical temperature. Indeed, as the temperature approaches
$T_{\rm c}^{\rm P}$, the enhanced order parameter $\Delta_P(T)$
approaches a finite (though small) value, $\Delta_P(T_{\rm c}^{\rm
P}-0)=(1/2) \hbar \omega$, and vanishes through a jump at
$T>T_{\rm c}^{\rm P}$ \cite{Ivlev,Klapwijk,Dmitriev2}, whereas the
critical current vanishes continuously, without any jump. Thus,
the temperature dependence of the critical current cannot in
principle be adequately described by equation of type
\eqref{Klap}; incidentally, this can be seen from a pronounced
deviation of the formula \eqref{Klap} from the experimental points
in a close vicinity of $T_{\rm c}$. In the present paper, the
experimental data will be analyzed by means of the exact formula
\eqref{IcP}, in which the numerical solution of equation
\eqref{EqDelta} for the quantity $\Delta_{\rm m}$ is used.

\section{Experimental results}

\begin{table}\vspace{-2mm}
\caption{Parameters of the film samples: $L$ is the length, $w$
the width, $d$ the thickness of the sample, and $l_{\rm i}$ is the
electron mean free path.} \label{tab}
\begin{center}
\begin{tabular}{|c|c|c|c|c|c|c|c|c|}
\hline
 Sample &    $L$, &   $w$,  &   $d$, &  $R_{4.2}$, &
$R^\square$, &
$T_{\rm c}$, & $l_{i}$, &  $R_{300}$, \\
       &  $\mu$m &  $\mu$m &   nm   &  $\Omega$   &  $\Omega$    &
K       &  nm      &  $\Omega$   \\
\hline
 SnW8   &  84     &   25    &    136 &  0.206      &  0.061
&
3.816  &   148   &   3.425 \\

SnW10  &  88     &   7.3   &    181 &  0.487      &  0.040       &
3.809  &   169   &   9.156 \\

SnW13  &  90     &   18    &    332 &  0.038      &  0.008       &
3.836  &   466   &   1.880 \\
\hline
\end{tabular}
\end{center}
\vspace{-6mm}
\end{table}

We investigate superconducting Sn thin films fabricated by a novel
technique \cite{Dmitriev} which ensures minimum defects both at
the film edge and in its bulk. The critical current of such
samples approaches the maximum possible theoretical value
\cite{Aslamazov}. This implies that the current density at the
film edges approaches a value of the order of $j_{\rm c}^{\rm GL}$
and thereby indicates the absence of the edge defects which might
produce local reduction of the edge barrier to the vortex entry
and corresponding decrease in $I_{\rm c}$. While measuring the
$I$-$V$ curves (IVC), the samples were placed in a double screen
of annealed permalloy. The $I$-$V$ curves were measured by a
four-probe method. The external irradiation was applied to the
sample which was placed inside a rectangular wave guide. The
electric component of the microwave field in the wave guide was
directed parallel to the transport current in the sample. The
parameters of some measured films are listed in the Table
\ref{tab}; conventional values $v_{\rm F}=6.5 \times 10^{7}$ cm/s
and $\tau _{\varepsilon}=4 \times 10^{-10}$ s were used for the
Fermi velocity and inelastic relaxation time of Sn.

\begin{figure}[htb]
\centerline{\includegraphics[height=3in]{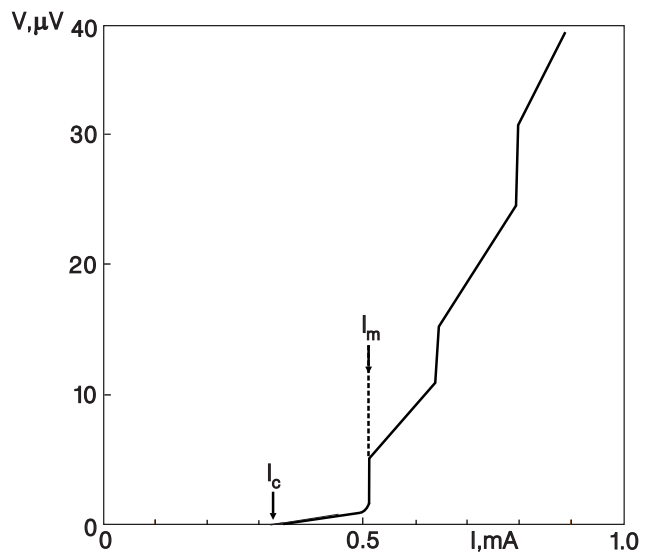}}\vspace{-2mm}
\caption{Typical $I$-$V$ characteristic of a wide [$w \gg \xi(T),
\lambda_\perp(T)$] superconducting film (sample SnW13) at the
temperature $T=3.798$~K. the temperature T=3.798 K. $I_{\rm m}$ is
the maximum current of the existence of the vortex state. $I_{\rm
c}$ is the critical current of the wide film.}
\label{f}\vspace{-3mm}
\end{figure}

The IVC of one of the samples is shown in figure \ref{f}. The film
resistivity caused by motion of the Abrikosov vortices occurs
within the current region $I_{\rm c}<I<I_{\rm m}$ (vortex portion
of the IVC), where $I_{\rm m}$ is the maximum current of the
existence of the vortex state \cite{Aslamazov,Dmitriev}. When the
current exceeds $I_{\rm m}$, the IVC shows voltage steps
indicating appearance of phase-slip lines.

\begin{figure}\vspace{-4mm}
\centerline{\includegraphics[height=3.2in]{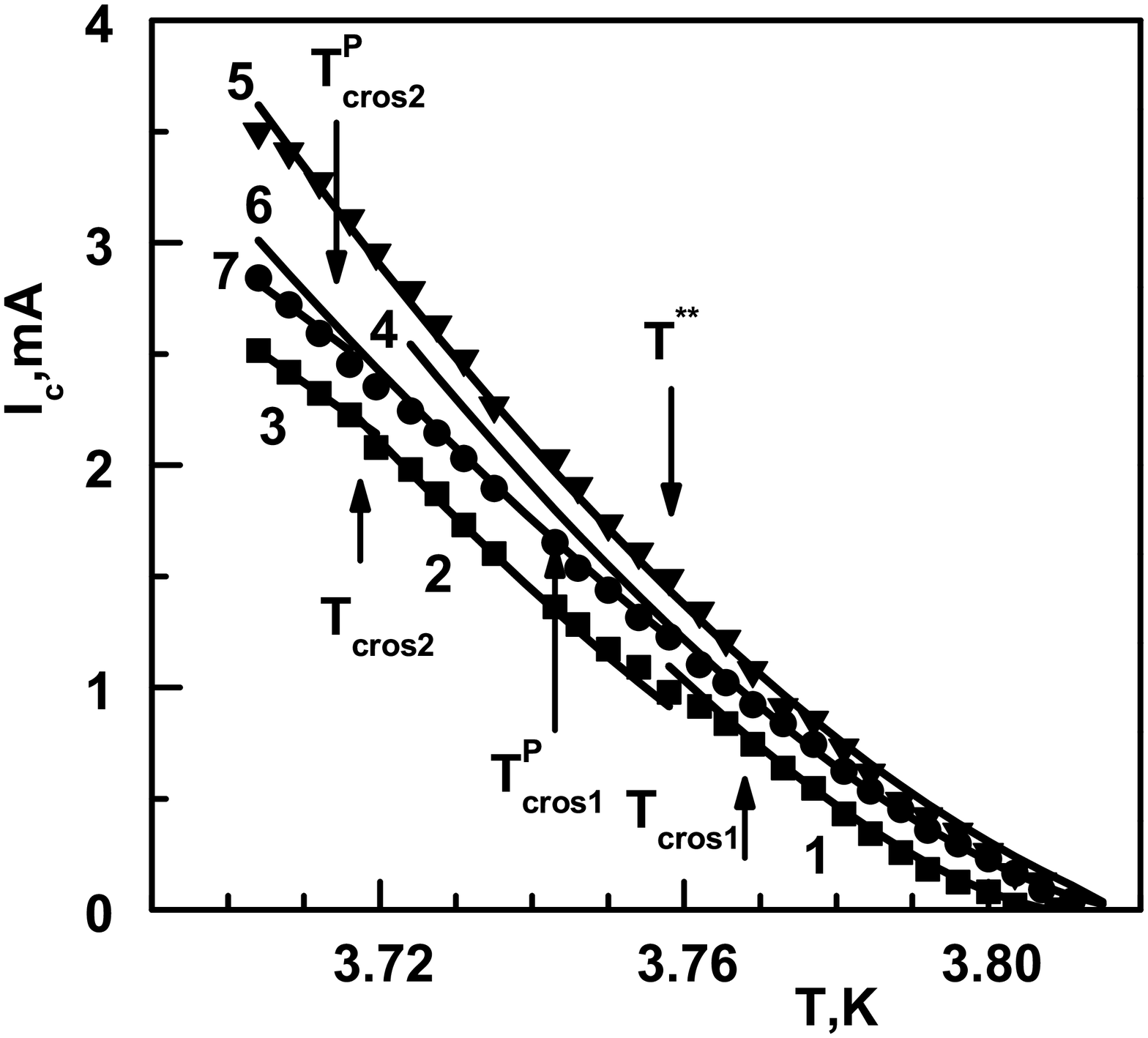}}\vspace{-2mm}
\caption{ Experimental temperature dependences of the critical
current for the sample SnW10 shown by symbols: $I_{\rm c}(P=0)$ --
$\protect\blacksquare$, $I_{\rm c}(f=9.2~\rm{GHz})$ --
$\protect\blacktriangle$, and $I_{\rm c}(f=12.9~\rm{GHz})$ --
$\protect\blacktriangledown$. Theoretical and
approximating dependences are shown by curves:\\
1 -- theoretical dependence $I^{\rm GL}_{\rm c}(T) = 7.07\times
10^2(1-T/T_{\rm c})^{3/2}$ mA [see \eqref{IcGL}]; \\
2 -- calculated dependence $I_{\rm c}(T)$ = 5.9$\times
10^2(1-T/T_{\rm c})^{3/2}$ mA; \\
3 -- linear theoretical dependence $I^{AL}_{\rm c}(T)$ =
9.12$\times
10^{1}(1-T/T_{\rm c})$ mA \cite{Aslamazov}; \\
4 -- theoretical dependence $I_{\rm c}(f=9.2~\rm{GHz})$ calculated
from \eqref{IcP}; this curve can be approximated by dependence
$I_{\rm c}(T) = 6.5\times
10^2(1- T/3.818)^{3/2}$ mA; \\
5 -- theoretical dependence $I_{\rm c}(f=12.9~\rm{GHz})$
calculated from \eqref{IcP}; this curve can be approximated by
dependence
$I_{\rm c}(T) = 6.7\times 10^2(1-T/3.822)^{3/2}$ mA;\\
6 -- theoretical dependence $I_{\rm c}(f=9.2~\rm{GHz})$ calculated
from \eqref{IcP} and normalized to curve 2; curve 6 can be
approximated by
dependence $I_{\rm c}(T)$ = 5.9$\times 10^2(1-T/3.818)^{3/2}$ mA; \\
7 -- linear approximating dependence $I_{\rm c}(T)$ = 9.4$\times
10^1(1-T/3.818)$ mA. \label{ff} }\vspace{-6mm}
\end{figure}

Figure \ref{ff} shows experimental temperature dependences of the
critical current for the sample SnW10. First we consider the
behaviour of $I_{\rm c}(T)$ with no electromagnetic field applied
(squares). The film width is rather small ($w=7.3~\mu$m),
therefore the sample behaves as a narrow channel within a
relatively wide temperature range, $T_{\rm cros1}<T<T_{\rm
c}=3.809$~K, in which the critical current is equal to the GL
pair-breaking current $I_{\rm c}^{\rm GL}(T) \propto (1-T/T_{\rm
c})^{3/2}$. A crossover temperature, $T_{\rm cros1}=3.769$~K,
corresponds to the transition to the wide film regime: at
$T<T_{\rm cros1}$, the vortex portion in the IVC occurs. The
temperature dependence $I_{\rm c}(T)$ at $T<T_{\rm cros1}$
initially holds the form $(1- T/T_{\rm c})^{3/2}$; then the value
of $I_{\rm c}(T)$ becomes smaller than $I_{\rm c}^{\rm GL}(T)$
below a certain characteristic temperature, $T^{**} \approx
3.76$~K, which is due to formation of inhomogeneous distribution
of the current density and its decrease away from the film edges.
Finally, at $T<T_{\rm cros2}=3.717$~K, the temperature dependence
of the critical current becomes linear, $I_{\rm c}(T)=I^{AL}_{\rm
c}(T)$ = 9.12$\times 10^1(1-T/T_{\rm c})$ mA, in accord with the
AL theory \cite{Aslamazov}.

In our measurements in a microwave field, the irradiation power
was selected to achieve a maximum critical current $I_{\rm c}^{\rm
P}(T)$. First we discuss the behaviour of $I_{\rm c}^{\rm P}(T)$
for the sample SnW10 in the microwave field of the frequency
$f=9.2$~GHz (figure \ref{ff}, triangles). In the temperature range
$T_{\rm cros1}^{\rm P}(9.2~{\rm{GHz}}) =3.744$~K~$<T<T_{\rm
c}^{\rm P}(9.2~{\rm{GHz}})=3.818$~K, no vortex portion in the IVC
was observed similar to the narrow channel case. We see that under
optimum enhancement, the narrow channel regime holds down to the
temperature $T_{\rm cros1}^{\rm P}(9.2~{\rm{GHz}})$ lower than
that in equilibrium state, $T_{\rm cros1}$. Furthermore, within
the region $T^{**} =3.760$~K$<T<T_{\rm c}^{\rm P}$, the
experimental values of $I_{\rm c}^{\rm P}$ are in good agreement
with those calculated from equation \eqref{IcP} (curve 4 in figure
\ref{ff}), in which the microwave power (quantity $\alpha$) was a
fitting parameter. Below the temperature $T^{**}$, the
experimental values of $I_{\rm c}^{\rm P}$ descend below the
theoretical curve 4, similar to the values of the equilibrium
critical current discussed above. Nevertheless, the experimental
points can be well fitted by equation \eqref{IcP} (figure
\ref{ff}, curve 6) normalized with a supplementary numerical
factor which provides agreement of this equation at zero microwave
field with the measured equilibrium critical current $I_{\rm
c}(T)$. We interpret this factor as the form-factor which takes
qualitative account of inhomogeneity of the current distribution
across the film width. Eventually, at $T<T_{\rm cros2}^{\rm
P}(9.2~\rm{GHz})=3.717$~K, the temperature dependence of the
critical current becomes linear (figure \ref{ff}, straight line
7).

The temperature dependence of the enhanced critical current
measured at higher frequency, $f=12.9~{\rm GHz}$, (figure
\ref{ff}, turned triangles) shows both the critical current and
the superconducting transition temperature, $T_{\rm c}^{\rm
P}(12.9~{\rm GHz})=3.822$~K~$>T_{\rm c}^{\rm P}(9.2~{\rm GHz})$,
to increase with the frequency, like in narrow channels.
Furthermore, at this frequency, the IVC shows no vortex portion in
the temperature range studied, down to $3.700$~K and even somewhat
below. Thus the temperature $T_{\rm cros1}^{\rm P}$ of the
transition to the wide film regime (not shown in figure \ref{ff})
decreases, $T_{\rm cros1}^{\rm P}(12.9~{\rm GHz})< 3.700~{\rm
K}<T_{\rm cros1}^{\rm P}(9.2~{\rm GHz})$ which considerably
extends the region of the narrow channel regime. It is interesting
to note that the experimental dependence $I_{\rm c}^{\rm P}(T)$ is
in good agreement with the equation \eqref{IcP} (curve 5) without
any additional normalization in the whole temperature range
studied. This would mean that in moderately wide films, the
temperature $T^{\ast \ast}$ of deviation of the experimental
dependence $I_{\rm c}^{\rm P}(T)$ from the Eliashberg theory is
likely to decrease at high enough frequency, similar to the
crossover temperature $T_{\rm cros1}^{\rm P}$.

\begin{figure}[htb]\vspace{-4mm}
\centerline{\includegraphics[height=3in]{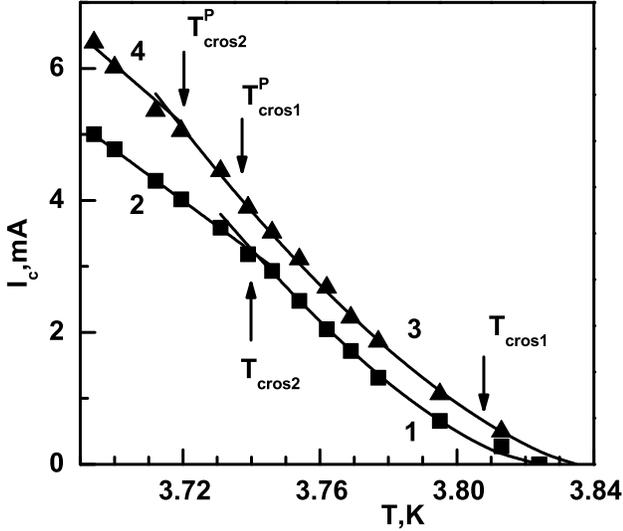}}\vspace{-2mm}
\caption{
Experimental temperature dependences of the critical current $I_{\rm
c}(P=0)$ -- $\blacksquare$ and  $I_{\rm c}(f=15.2~{\rm GHz})$
-- $\blacktriangle$, for the sample SnW8.\\
Theoretical and
approximating dependences are shown by curves:\\
1 -- calculated dependence $I_{\rm c}(T)$ = 1.0$\times
10^{3}(1-T/T_{\rm c})^{3/2}$ mA;\\
2 -- linear theoretical dependence $I^{AL}_{\rm c}(T)$ =
1.47$\times 10^2(1-T/T_{\rm c})$ mA \cite{Aslamazov}; \\
3 -- theoretical dependence $I_{\rm c}(f=15.2 GHz)$ calculated by
\eqref{IcP} and normalized to curve 1, curve 3 can be approximated
by dependence $I_{\rm c}^{\rm P}(T)$ = 1.0$\times
10^{3}(1-T/3.835)^{3/2}$ mA;\\
4 -- linear calculated dependence $I_{\rm c}^{\rm P}(T)$ =
1.72$\times 10^2(1-T/3.835)$ mA. \label{fff} } \vspace{-4mm}
\end{figure}

The temperature dependences of the critical current for the
sample SnW8 are shown in figure \ref{fff}. We begin with the
analysis of the behaviour of $I_{\rm c}(T)$ with no
electromagnetic field applied. The film width is relatively large,
$w$=25~$\mu$m, therefore this sample can be treated as a narrow
channel only very close to $T_{\rm c} = 3.816$~K; at
$T<T_{\rm cros1}=3.808$~K it behaves as a wide film, with the
vortex portion in the IVC. The temperature dependence of the
critical current holds the form $(1-T/T_{\rm c})^{3/2}$ from
$T_{\rm c}$ down to $T_{\rm cros2}=3.740$~K, though the value of
$I_{\rm c}$ is smaller than $I^{\rm GL}_{\rm c}$ within this
temperature range. This means that substantial current
inhomogeneity develops very close to $T_{\rm c}$ as well (the
difference between $T_{\rm c}$ and the characteristic temperature
$T^{**}$ cannot be reliably resolved). At $T<T_{\rm cros2}$, the
temperature dependence of the critical current becomes linear,
according to the AL theory \cite{Aslamazov}: $I_{\rm c}(T)$ =
$I^{AL}_{\rm c}(T)$ = 1.47$\times 10^2(1-T/T_{\rm c})$ mA.

In the microwave field of the frequency $f=15.2~{\rm GHz}$, the
superconducting transition temperature increases up to $T_{\rm
c}^{\rm P}=3.835$~K, whereas the temperatures of both the
crossover to the linear AL dependence, $T_{\rm cros2}^{\rm
P}=3.720~$K, and the transition to the wide film regime, $T_{\rm
cros1}^{\rm P} = 3.738$~K, exhibit noticeable decrease. At the
same time, in order to achieve good agreement between the
experimental dependence $I_{\rm c}^{\rm P}(T)$ and equation
\eqref{IcP}, we must apply the normalization of this equation on
the measured equilibrium critical current $I_{\rm c}(T) =
1.0\times 10^{3}(1-T/T_{\rm c})^{3/2}$ mA within the whole
temperature range $T>T_{\rm cros2}^{\rm P}$ (figure \ref{fff},
curve 3) including the close vicinity of the critical temperature
(the temperature $T^{**}$ is still indistinguishable from $T_{\rm
c}^{\rm P}$). Thus, in rather wide films, even the frequency
$f=15.2$~GHz is not high enough to transform the absolute
magnitude of $I_{\rm c}^{\rm P}(T)$ to the bare dependence
\eqref{IcP} calculated for a narrow channel, unlike the case of
the relatively narrow sample SnW10 at $f=12.9$~GHz discussed
above.

\section{Discussion of the results}

We begin this section with discussion of the superconductivity
enhancement effect which manifests itself as increase in the
superconducting transition temperature $T_{\rm c}^{\rm P}$ and in
the magnitude of the critical current $I_{\rm c}^{\rm P}$ compared
with their equilibrium values. Qualitative similarity of the
results with those obtained for the narrow channels
\cite{Dmitriev1} and the possibility to quantitatively describe
the temperature dependence $I_{\rm c}^{\rm P}$ by the equations of
the Eliashberg theory convince us that the mechanism of the
enhancement effect is the same for the wide films and the narrow
channels, i.e., it consists in enhancement of the energy gap
caused by a redistribution of the microwave excited nonequilibrium
quasiparticles to higher energies \cite{Eliashberg1}.

This conclusion seems to be not quite evident for the wide films
with inhomogeneous current distribution across the sample because
it is the current inhomogeneity that may be suggested to
completely suppress the superconductivity stimulation in bulk
superconductors. Indeed, in the latter case, the concentration of
the transport current and the microwave field within the Meissner
layer near the metal surface gives rise to an extra mechanism of
quasiparticle relaxation, namely, spatial diffusion of the
microwave excited nonequilibrium quasiparticles from the surface
to the equilibrium bulk. The efficiency of this mechanism is
determined by the time of quasiparticle escape, $ \tau_{\rm
D}(T)=\lambda^2(T)/D$, from the Meissner layer, which is
three-to-four orders of magnitude smaller than typical inelastic
relaxation time. Such high efficiency of the diffusion relaxation
mechanism is likely to result in the suppression of the
enhancement effect in bulk superconductors. However, relying upon
a qualitative difference between the current states in bulk and
thin-film superconductors outlined in Introduction, one can argue
that moderate non-uniformity of the current distribution in wide
films (with no concentration of an exciting factor at short
distances) causes no fatal consequences for the enhancement
effect, and that the diffusion of nonequilibrium quasiparticles
excited within the whole bulk of the film introduces only
insignificant quantitative deviations from the Eliashberg theory.

In our consideration, we used a model approach to take these
deviations into account by introducing the numerical form-factor
of the current distribution into the formula \eqref{IcP} for the
enhanced critical current. We evaluate this form-factor by fitting
the limit form of equation \eqref{IcP} at zero microwave power,
i.e., equation \eqref{IcGL}, to the measured values of the
equilibrium critical current. Then we apply obtained values of the
form-factor 0.83 for SnW10 and 0.57 for SnW8 to equation
\eqref{IcP} at $P\neq 0$, which results in a considerably good fit
to the experimental data, as demonstrated in previous section.

There is a question worth to discuss how the Eliashberg
mechanism works in the regime of a wide film, i.e., when the
superconductivity breaking is due to the vortex nucleation rather
than due to exceeding the maximum value allowed by the
pair-breaking curve by the transport current. We believe that in
this case the enhancement of the energy gap results in
corresponding growth of the edge barrier for the vortex entry, and
this is what enhances the critical current in the wide film
regime. It is interesting to note that no essential features appear
in the curves $I_{\rm c}(T)$ when the films enter the
vortex resistivity regime. From this we conclude that the
transition between the regimes of the uniform pair-breaking and
vortex resistivity has no effect on both the magnitude and the
temperature dependence of the critical current.

To complete with the discussion of the superconductivity
enhancement effect, we draw one's attention to the empiric fact
that all the fitting curves for $I_{\rm c}^{\rm P}(T)$ obtained by
equations of the Eliashberg theory are excellently approximated by
the power law $(1-T/T_{\rm c}^{\rm P})^{3/2}$. This law is quite
similar to the temperature dependence of the GL pair-breaking
current, in which the critical temperature $T_{\rm c}$ is replaced
by its enhanced value $T_{\rm c}^{\rm P}$. Explicit expressions
for such approximating dependencies, with numerical coefficients,
are presented in captions to the figures \ref{ff} and \ref{fff}.

The next important result of our studies is the essential
extension of the temperature range of the narrow channel regime of
wide films on enhancement of superconductivity: in the microwave
field, the temperature of the crossover to the wide film regime,
$T_{\rm cros1}^{\rm P}$, considerably decreases as compared with
its equilibrium value, $T_{\rm cros1}$. In the first glance, this
result somewhat contradicts the criterion of the transition
between the different regimes mentioned in Introduction, $w =
4\lambda_\perp(T_{\rm cros1})$, because increasing energy gap under
irradiation implies decreasing magnitude of $\lambda_\perp$ and,
correspondingly, decreasing characteristic size of the vortices.
This obviously facilitates the conditions of the vortex entry to
the film, thus the crossover temperature should {\em increase} in
the microwave field. We believe, however, that there exists a more
powerful stabilizing effect of irradiation on the vortices. The
role of this effect is to delay the vortex nucleation and/or
motion and to maintain the existence of the narrow channel regime
down to low enough temperatures at which the current inhomogeneity
eventually becomes well developed. Indeed, for the sample SnW10 at
zero microwave power, the transition to the wide film regime
occurs at the temperature $T_{\rm cros1}^{\rm P}(T)$ {\em higher}
than the temperature $T^{**}$ at which the deviation of $I_{\rm
c}^{\rm P}(T)$ from the GL uniform pair-breaking current begins to
be observed. This means that in equilibrium case the vortex
nucleation begins at relatively small inhomogeneity which still
weakly affects the magnitude of the critical current. In contrast
to this, under microwave irradiation of the frequency $f=9.2$~GHz,
the vortex resistivity occurs at the temperature $T_{\rm
cros1}^{\rm P}$ {\em lower} than $T^{**}$, i.e., the inhomogeneous
current state shows enhanced stability with respect to the vortex
nucleation in the presence of a microwave power. Similar
conclusion can be drawn regarding the behaviour of the critical
current in the wider film SnW8.

We suggest the following qualitative explanation of the
stabilization effect. Since the dimensions of the samples are
small compared to the electromagnetic wavelength (the sample
length is $\sim 10^{-4}$ m and the minimum wavelength is $\sim
10^{-2}$ m), we deal, in fact, with an alternating high-frequency
current, $I_{\rm f} \propto \sqrt{P}$, flowing through the sample.
The relative power $P/P_{\rm c}\sim 0.1 \div 0.2$, corresponding
to the maximum enhanced current $I_{\rm c}^{\rm P}(T)$, is rather
high, therefore the amplitude of $I_{\rm f}$ may be comparable
with the magnitude of the critical transport current. This results
in a considerable modulation of the net current flow through the
sample which presumably enhances the stability of the current flow
with respect to the nucleation and motion of the vortices. A
possible analogy of such phenomenon is given by stabilization of
the current state in narrow channels at supercritical currents
caused by self-radiation of phase-slip centers \cite{PSC}.

As noted in previous Section, the stabilizing effect becomes more
pronounced with increasing irradiation frequency $f$: the region
of the narrow channel regime considerably extends as the frequency
grows. In the framework of our assumption about the stabilizing role
of the high-frequency modulation of the current flow, such effect
can be explained as follows. The relative microwave power of
optimum enhancement, $P/P_{\rm c}$, was found to increase with $f$
\cite{Dmitriev1}, while the critical power $P_{\rm c}$ changes
with $f$ only at small enough frequencies, $f< f_\Delta$
\cite{Pals,Bezuglyi}, where $f_{\Delta} \approx (1-T/T_{\rm
c})^{1/2} / 2.4 \tau_{\varepsilon}$ is the inverse of the gap
relaxation time and does not exceed $0.1$~GHz for our samples and
temperatures. At larger frequencies used in our experiments,
$P_{\rm c}$ holds a constant value, which enables us to attribute
the variations in $P/P_{\rm c}$  to variations in the absolute
magnitude of the irradiation power $P$, i.e., in the amplitude of
the modulating high-frequency current $I_{\rm f}$. Thus the
increase in the irradiation frequency causes the microwave power
of optimum enhancement of superconductivity to increase and,
correspondingly, the stabilizing effect of the electromagnetic
field on the vortices to enhance.

\section{Conclusions}

The results of our investigation enable us to conclude that the
mechanism of superconductivity enhancement by a microwave field is
the same for both the narrow channels and wide films -- this is
the Eliashberg mechanism of enhancement of the superconducting energy gap
caused by the excitation of nonequilibrium quasiparticles to a
high energy region within the whole bulk of a superconductor. In
the vicinity of $T_{\rm c}$, where any film can be treated as
a narrow channel, the Eliashberg mechanism enhances the
critical current in a usual way, through modification of the
pair-breaking curve for the superconducting current. At lower
temperatures, when the wide film enters the regime of the vortex
resistivity, the enhancement of the critical current is likely
associated with the growth of the edge barrier to the vortex
entry, nevertheless giving the magnitude and the temperature
dependence of the critical current quite similar to that in the
previous case. Such similarity extends down to low enough
temperatures, at which a linear temperature dependence predicted
for extremely wide films is observed.

Another important effect of the microwave irradiation is to
stabilize the current state of the film in respect to the entry of
the vortices, presumably by deep modulation of the transport
current by the induced high-frequency current. This results in the
extension of the temperature range in which the film behaves as a
narrow channel, showing no vortex resistivity in the
current-voltage characteristic. The stabilizing effect grows with
frequency which is explained by simultaneous increase in the
pumping power and, correspondingly, in the amplitude of the
high-frequency current.

We achieve good accordance of the experimental data with Eliashberg theory by
introducing a numerical form-factor into equations initially
derived for homogeneous distribution. Despite the simplicity and
high enough quality of such approximation, this problem requires a
more consistent approach, involving solution of diffusive
equations of nonequilibrium superconductivity in a spatially
inhomogeneous case.

\section{Acknowledgments}
The authors are grateful to T.V.Salenkova for preparing the films
and to E.V.Khristenko for helpful discussions.

\end{document}